\documentclass[conference, a4paper, 10pt]{IEEEtran}
\usepackage{blindtext, graphicx, amsmath}
\usepackage[font=footnotesize]{caption}
\usepackage{enumitem, comment, multirow}
\usepackage[usenames, dvipsnames]{color}
\graphicspath{ {images/} }
\ifCLASSINFOpdf
\else
\fi
\let\oldIEEEkeywords\IEEEkeywords
\def\IEEEkeywords{\oldIEEEkeywords\itshape}

\begin{document}
\title{Effective Predictions of Gaokao Admission Scores for College Applications in Mainland China}
% University Score Prediction and Recommendation System}

\author{
\IEEEauthorblockN{Hao Zhang}
\IEEEauthorblockA{
Department of Computer Science\\
University of Massachusetts\\
%\\Computer Science\\
Lowell, MA\\}
\and
\IEEEauthorblockN{Jie Wang}
\IEEEauthorblockA{
Department of Computer Science\\
University of Massachusetts\\
%\\Computer Science\\
Lowell, MA\\}
%\and
%\IEEEauthorblockN{James Kirk\\ and Montgomery Scott}
%\IEEEauthorblockA{Starfleet Academy\\
%San Francisco, California 96678-2391\\
%Fax: (888) 555--1212}
}


%Telephone: (800) 555--1212, Fax: (888) 555--1212}


\maketitle

\begin{abstract}
Gaokao is the annual academic qualification examination for college admissions in mainland China. Organized by each provincial-level administrative region (PAR), Gaokao takes place at the same time nationwide in early June. To enroll in a university in September, students
must take Gaokao and submit common applications for admission to their home PAR Gaokao office in July, listing a small and fixed number of universities and majors they intend to attend and study. About 9.5 million high-school seniors participate in Gaokao every year, and the Gaokao scores are good for just one year.
A student has a strong chance to be accepted
%which is likely to happen
if their Gaokao score is better than the admission scores of the universities they selected in their applications. However, %the admission scores of universities and their majors are
the admission scores of universities are unknown at the time when filling out applications, which to be determined dynamically during the admission process and may fluctuate from year to year.
%
%match with their Gaokao scores and intention. It is notable
To increase their chances of acceptance to a best-suited university,
students need to
%select universities and majors wisely.
%Since nobody knows the admission scores to any university or %major at the time of filing out applications, this becomes a %guessing game.
%
%that applicants must make their decisions in limited time, with %limited knowledge about the university admission score and %the university itself.
%Thus, providing
predict %predictions of
admission score of each university they are interested in.
% has thus attracted much attention.
%
%Each year many people and a number of companies offer %predictions of admission scores for a university and its majors.
Early prediction methods are empirical without the backing of in-depth data studies.
We fill this void by presenting well-tested mathematical models based on the ranking of Gaokao scores in a PAR. %and
%validations through extensive experiments..
%We present four methods for university admission score %prediction and an automated recommendation system to help %applicants make proper decisions.
We show that our methods significantly outperform the
methods commonly used by teachers and experts, and
can
predict admission scores with an accuracy of about 91\% within a 7-point margin in an exam of a 750-point grading scale.
\end{abstract}


% \fi
\IEEEpeerreviewmaketitle

\section{Introduction}
Almost all people who plan to pursue higher education in mainland China must take the Gaokao exams for
college admissions.
%Qfor the most part, depend only on students' Gaokao scores.
There are two types of exams: the ``Li-Ke" exam (meaning the science exam) and the ``Wen-Ke" exam (meaning the liberal-arts exam).
Students who are interested in pursuing a major in science and engineering are required to take the ``Li-Ke" exam, and students who are interested in pursuing a major in art and humanity and social sciences are required to take the ``Wen-Ke" exam.
%College admissions follow the following procedure.
Depending on students' Gaokao scores and the admission quota for each PAR, the provincial Gaokao office determines an appropriate minimum admission score for each tier of universities. This score is the provincial Admission Score Cutoff Line (ASCL).
The total of about 2,400 universities and colleges in mainland China are officially designated into three tiers based on the qualities of the degree programs they offered.
After the announcement of ASCL, students only have a few days to submit their applications. They are allowed to apply for universities of a particular tier only if they pass the corresponding ASCL for that tier. Early policies such as ``apply before taking exams'' and  ``apply after taking exams but before knowing scores'' were eliminated in 2017. Recently, some PARs have re-designated universities of three tiers into only two tiers, and this seems to be
the current trends.

To guarantee that each student is admitted by at most one university, each PAR uses one of the following models
to restrict how admission staff members can access applications: parallel admission, gradient admission, and hybrid admission.
The first model follows the principle of ``scores first, according to preference'' and the
second model follows the principle of ``preference first, according to scores.''
In recent years, increasingly more PARs have adopted the parallel admission model.

Under any of these admission models,
students need to guess the admission scores
of a university and its majors to obtain
admissions that better match their abilities and interests.

%
%When making their decisions on which universities and majors %to apply, they don't know the admission scores of any %university and its majors.
Without a trusted prediction model, what students can rely on is the following: published ASCLs of their home PAR; inputs from teachers and other sources; and the statistics of admission scores in the past, for example, each university's
minimum, maximum, average admission scores, and the number
of admitted students. Some PARs may also provide such statistics down to each major for each university. This process, unfortunately, is prone to substantially overestimating certain universities' admission scores while underestimating the others.
%
%Second, applicants do not have much knowledge about universities,
%such as rankings of the college and major, employment records,
%employment trends, locations, and tuitions.
%are important factors should be taken into consideration when %making the decision. However, most applicants have little-known %about these information and collecting the data is time-consuming.
%	Finally,
An added difficulty is that applicants only have a few days to make decisions (see
%Figure \ref{fig:1}
Table \ref{table:1}).
\begin{table}[b]
\footnotesize
\centering
\caption{The distribution of the number of days given to students to fill out applications by each province in 2017}\label{table:1}
\begin{tabular}{l|c|c|c|c|c|c|c|c}
\hline
\textbf{Days} & 4 & 5 & 6 & 7 & 8 & 9 & 10 & $>10$ \\\hline
\textbf{PARs} & 2 & 8 & 4 & 4 & 4 &3 & 2 & 4\\\hline
\end{tabular}
\end{table}
For example,
%
%According to our survey,
in 2017, students in 18 PARs (there are a total of 33 PARs in mainland China)
only have at most seven days to submit their applications (some only have four days)
%and
%about 60\% of provinces require applicants to submit their %applications within seven days
after they are notified of their Gaokao scores
shortly after
the ASCLs are published.
Because of overestimates or underestimates of admission scores of the universities and majors they apply, applicants may end up being admitted
by a university and major that poorly match their intention and ability or
receiving no offers at all even though they could be admitted by some
universities,
resulting in mismatching admissions time and time again.

\begin{comment}
    \begin{figure}[h]
    \centering
    \vspace{-0.45cm}
    \captionsetup{justification=centering}
    \includegraphics[width=8cm, height=6cm]{time.png}
    \caption{The distribution of days given to students to fill out applications by each province in 2017} \label{fig:1}
	\end{figure}
\end{comment}

Despite its importance,
early efforts on predicting admission scores have
not produced satisfactory models. Most prediction methods
published so far
are based on past admission records with straightforward statistics.
We attempt to fill this gap by analyzing extensive data sets of Gaokao admission records down to each student's Gaokao
score and which university and major that student was admitted
to (with student's identification removed).
Our major contributions are described below:

\begin{enumerate}
\item We present a series of mathematical prediction models
based on the rankings of Gaokao scores. Our experiments
show that our models can
predict admission scores with an accuracy of about 91\% within a 7-point margin in an exam of a 750-point grading scale.

\item We modify an existing recommendation system
using General Morphological Analysis (GMA) to
provide reliable recommendations to students \cite{Lu2016}
according to the parallel admission model.
\end{enumerate}

This paper is organized as follows. In Section 2, we describe
existing methods for predicting admission scores. We then
present in Section 3 our ``ranking-based'' prediction models.
 In Section 4, we present a recommendation system using General Morphological Analysis (GMA) based on predicted admission scores, which is a modification of our early recommendation system \cite{Lu2016}. In section 5, we will describe our evaluating datasets, data pre-processing methods, and prediction accuracy comparisons of our methods and the standard methods used by most people.

\section{Related work}

It is a common perception that knowing universities' admission scores is the most important factor in
determining how students should fill out applications for admissions. This perception was confirmed by
Zhao and Wang \cite{Zhao2012}.
%
%
% influencing Gaokao acceptance (Zhao et al., 2012), university %admission score is the most important one, even more than %Gaokao score.
%Thus, a fast and reliable prediction of admission scores is the %key to address the limitations.
%Thus, an accurate prediction result will help reduce the
%risk of mismatch.
%applying risk and increase the hit ratio, but it also is a %foundation for a recommendation system.
%
%Although university admission scores are widely considered i%mportant, little work has been done in predicting admission scores. %For many Gaokao models and applications, the prediction model is %weak or even does not exist.
%add the result for "gaokao application report"

All existing score-prediction methods used by high-school teachers and experts are
empirical, and all methods published so far only exploit past admission scores. We
refer to these methods as ``scoring methods.''
For example, Li et al. \cite{Li2010} assumed that the average admission score of each university is stable with a small variation each year, regardless of the ASCL scores.
We refer to this model as the average admission score (AAS) model.
%
%in the past years is
%the baseline in most of these methods.
%
%One common way is using , including a parameter that represents the %student's risk appetite (Li et al., 2010). This method assumes the %university admission score is stable with a small variation each year, %regardless of the PAD change.

Zhao and Wang \cite{Zhao2012} and Zhu et al. \cite{Zhu2006}
assumed that the differences between the PAR ACSLs and
the admission scores of each university over the years are
stable with only minor variations. We refer to this model as
the average admission difference (AAD) model.
%
%provincial and university admission score is stable, predicts score by %adding the average admission difference (AAD) to current year PAD %%(Zhao et al., 2012 \& Zhu et al., 2006).

However, Wang and Wu \cite{Wang2017} showed that for the past ten years, the differences between the ASCLs and the admission scores of
certain tier-1 universities are consistently increasing. In particular, for top-ranking universities, the average differences have increased 45.9 points for the Li-Ke exam and 24.3 points for the Wen-Ke exam. %Therefore the stable admission difference assumption is questionable.

\section{Prediction models}

%We present
% four models to predict the university admission score, using past data.
We first describe our baseline model and then present
two improvements, each addressing a weakness in the baseline.

The underlying principle of our designs is the following observation:
Each university would likely admit students
of the same quality from the
same PAR, even though the ASCLs and the Gaokao scores of students of the same level in that PAR may fluctuate from year to year. This is particularly true
for universities and majors that are better established.
The ranking of a student's
Gaokao scores in the same PAR
would be a better indication of the student's quality, not the Gaokao score itself. For example, a university
would typically admit students whose Gaokao scores are of roughly the same ranks from the same PAR in the recent years,
for the quality of a university and the general perception of its prestige
does not change dramatically.

%Among all of our models, the ranking is one of the most %important factors.
%Since the admission score can be influenced by lots of %factors, such as admission policy, exam difficulty, etc., the %ranking will not.

%We will explain in detail in Section four.
In this paper, we will only focus on the prediction of the admission Gaokao score for each university, rather than the admission score for each major of a university.
Without loss of generality, in what follows our discussions are for a
particular PAR and we will not specifically mention which one. We assume that we have a score-to-ranking table for all students in a particular year,
where each Gaokao score corresponds to a ranking.
Let~$S_\ell(ranking)$~and~$R_\ell(score)$~denote, respectively, two score-ranking  operations, where $\ell$ represents the current year (e.g., $\ell = 2018$ this year), $S_\ell$ takes a ranking of a Gaokao score
in Year $\ell$  as input and finds the corresponding score using a score-ranking table in Year~$\ell$, while $R_\ell$~is the inverse operation of~$S_\ell$, which takes a Gaokao score as input and returns its ranking.

Note that for each Gaokao score $s$ there may be a large number of
students with that score. Thus, we define $R_\ell(s)$ to be the smallest ranking after all students
whose Gaokao scores are greater than $s$. In other words,
if there are $k$ students whose Gaokao scores are greater than $s$,
then $R_\ell(s) = k+1$.
Note that
%the number of students tied forthe same rank
%$|R_\ell(s) - R_\ell(s-1)|$ can
there may be
many students tied for the same rank,  as many as thousands, since on average each PAR has about 300 thousand students participating in Gaokao.

We will also use the following notations:
\begin{itemize}
\item $U$ denotes a university.
\item $A_\ell(U)$ denotes the admission score of $U$ in Year $\ell$.
\item $L_\ell$ denotes the ASCL in Year $\ell$ for universities of a particular tier,
which is also the lowest Gaokao score in Year $\ell$ for this admission tier; when
we need to specify which tier, we may use $L^j_\ell$ to denote the ASCL in Year $\ell$ for
universities of tier-$j$, where $j \in \{1,2,3\}$.
\item $N_\ell$ denotes the total number of students admitted in Year $\ell$ for
universities of a particular tier.
\item $H_\ell$ denotes the highest Gaokao score in Year $\ell$ for universities of a particular tier.
%\item $l_\ell$ denotes the Lowest Gaokao score in Year $\ell$ for universities of a particular tier.
\end{itemize}
%Let $u$ denote a university, ,
%and , "L" for "Lowest Score",
We use SRT to denote the score-ranking table.

% for "Highest Score".
%for "University Admission Score", "PAD" for "Provincial Admission Score", "H" for "Highest %Score", "L" for "Lowest Score", "SRT" for "Score-Ranking Table".

\subsection{Baseline ranking model (BRM)}

Our baseline model is based on the following assumption:
Each university admits students whose Gaokao scores are in
a stable range of rankings over the recent two to three years,
even though the Gaokao scores of the same rankings
may fluctuate. This is a reasonable assumption, particularly for
well-established universities with stable reputations.
%has its stable ranking interval when selecting applicants, the university admission score %might vary a lot, but the variation for the admitted students' ranking is relatively small.
Thus, assuming that the ranking of the admission scores for a university in two
consecutive years are the same, we have
\begin{equation} \label{eq:1}
    R_{\ell}(A_{\ell}(U)) = R_{\ell-1}(A_{\ell-1}(U)).
\end{equation}
Recall that $S_\ell$ is the inverse operation of $R_\ell$.
Thus,
%Therefore,
we can predict the admission score of $U$ in Year $\ell$ by applying $S_{\ell}$  on both sides of the equation as follows:
\begin{equation} \label{eq:2}
	A_{\ell}(U) = S_{\ell}(R_{\ell-1}(A_{\ell-1}(U))).
\end{equation}
%This model is referred to as the baseline ranking model (BRM).

\subsection{Weight slicing model (WSM)}

We note that BRM would work well
if the total number of students $N_{\ell}$ to be admitted in Year $\ell$
is about the same as the total number of students $N_{\ell-1}$ admitted in the previous
year $\ell-1$.
However, $N_{\ell}$ is unknown when students are filling out applications.
We may
instead look at
the difference of the ASCLs of Year $\ell$ and Year $\ell-1$
for a given tier of universities. Let
$$W = |L_{\ell} - L_{\ell-1}|.$$
%and analyzing news which usually would release some information whether
%the number of students to be admitted in Year $i+1$ would increase or %decrease over %Year $\ell$, or about the same.
Our experiments show that, when $W$ is small, the baseline ranking model performs well;
but when $W$ is large, predictions are less than satisfactory.

To resolve this problem, let us think of $W$ as weight
and divide it into different portions. We then distribute
each portion to a range of ranking proportionally.
%For convenience, let $ASCL_\ell^j$
%denote the ASCL of PAR $p$ in Year $i$
%for universities of tier-$j$, where $j = 1, 2, 3$.

In particular, we divide $W$ into $1, 2, \cdots, d$ such
that $d$ is the smallest integer satisfying the following
inequality:
$$\frac{d(d+1)}{2} \geq W.$$

Let $N_s$ denote the number of students whose Gaokao score is $s$.
Note that $N_s$ is typically much larger than $N_{s'}$ if $s' < s$.
%there are more students with the same lower Gaokao score.
That is,
the number of students at the same rank is larger if the corresponding Gaokao score is smaller.
We perform the following:
\begin{enumerate}
\item Recall that $N=N_\ell$ is the total number of universities of a particular tier in Year $\ell$. Sort these universities in descending order based on their admission scores as $U_1, U_2, \ldots, U_N$, where $U_1$ has the highest admission score and $U_N$ the lowest. If two universities have the same admission scores, we use out-of-band data to determine their ordering,  such as the
general perception of prestige, locations of the universities, among other things.
Let $$b_j = \frac{2(d-j)N}{d(d+1)}$$ for
		$j = 0, 1, \ldots d-1$.
It is easy to show that
$$\sum_{j=0}^{d-1}  b_j = N.$$

Let $b_j = \lfloor b'_j \rfloor$. We separate the interval $[1,N]$ into $d$ consecutive intervals according to their
indexes:
		$$[1, b'_0], [b'_0+1, b'_1+b_0], \ldots, \left[\sum_{j=0}^{d-2}b'_j + 1, N\right].$$

\item For a given university $U$ in the $j$-th interval, add $j$ to the baseline
		ranking model to make the prediction:
		\[
		A_{\ell}(U) =
		\begin{cases}
		S_{\ell}(R_{\ell-1}(A_{\ell-1}(U))) + j, & \mbox{if~$L_{\ell} \geq L_{\ell-1},$}\\
		S_{\ell}(R_{\ell-1}(A_{\ell-1}(U))) - j, & \mbox{otherwise.}\\
		\end{cases}
		\]
	\end{enumerate}
%We call this model weight slicing model (WSM).

\subsection{Weighted point model (WPM)}

Similar to WSM, the weighted point model also modifies the baseline
ranking model by adding a bias to the admission score of each university, but from a different perspective. The purpose is
to fix a problem resulted from the following scenario:
%We consider the following scenario.
Suppose that $L_{\ell}$ is much larger than $L_{\ell-1}$ and the total number of applicants
in Year $\ell$ is about the same as that in Year $\ell-1$. This implies that
% relatively small,
% the average number of students per point will increase, and vice versa.
the same Gaokao score that did not pass the admission score of university $U$ in the previous year may pass it this year.
In other words, more students have higher Gaokao scores than the previous year.
Since by assumption that
each university's admission quota stays about the same this year,
%since the university's enrollment capacity is a fixed number. Thus,
it has to increase its admission score accordingly.
%to maintain the capacity.
WPM simulates this process as follow:

\begin{enumerate}
		\item Let $\Delta$ denote the average number of students per point in the previous year. That is,
			$$\Delta = \left\lceil \frac{N_\ell}{H_{\ell-1} - L_{\ell-1}}\right\rceil.$$
			%where $R(L_{i} -1) - 1$ is the number of students who passed $L_\ell$.
		\item For a given university $U$, let~$D_\ell(U)$~denote the
			total number of students whose Gaokao score is exactly $A_\ell(U)$.
			%is one-point below $U$'s admission score.
			That is,
%
%			ranking difference between the admission score and one point below the score in %previous year.
			$$D_{\ell-1}(U) = |R_{\ell-1}(A_{\ell-1}(U)) - R_{\ell-1}(A_{\ell-1}(U)-1)|.$$
			\item Let~$j = 0$ if $D_{\ell-1}(U) < \Delta$. Otherwise, let~$j$~denote the smallest integer that satisfies the following inequality:
			$$j\Delta \leq D_{\ell-1}(Ui) < (j+1)\Delta,$$
			where $j = 1,2,\ldots, \lceil D_{\ell-1}(U)/\Delta\rceil.$
			\item For a given university~$U$, add $j$ to the baseline ranking model as follows:
			\[A_{\ell}(U) =
			\begin{cases}
			S_{\ell}(R_{\ell-1}(A_{\ell-1}(U))) + j, & \mbox{if~$L_{\ell} \geq L_{\ell-1}$,}\\
			S_{\ell}(R_{\ell-1}(A_{\ell-1}(U))) - j, & \mbox{otherwise.}\\
			\end{cases}
			\]
		\end{enumerate}

\section{Recommendation model}

We present a recommendation system that modifies our previous Gaokao application
recommendation
model called EEZY (meaning ``personalized recommendations") using
general morphological analysis (GMA) \cite{Lu2016}. Using a large volume of complete Gaokao admission data we collected over the past several years, our system can help students make decisions based on their Gaokao scores and interests and give them a matching measurement for each recommendation slot.
By ``complete Gaokao admission data'' it means for each year, the data consists of information $(S,U,M)$ of each admitted student, where $S$ is the student's Gaokao score,
$U$ and $M$ are the university and major that admit the student.
Without using a prediction model, however, EEZY could only match the applicant's Gaokao score with the previous year's data. We modify this system by adding a prediction model.

We define seven parameters for students to identify their personal preference, and another set of seven parameters for each university to represent its status (see Table \ref{table2}). The meanings of these parameters are self-explanatory.

	\begin{table}[b]
	\footnotesize
	\centering
	\caption{Parameters for EEZY}\label{table2}
		\begin{tabular}{l | l}
		\hline
		\textbf{Student preference} & \textbf{University status} \\
		\hline
		Gaokao Score       & Admission Score   \\
		Exam Type          & Major Type        \\
		Tier               & Admission Tier    \\
		Preferred Location & Location          \\
		Disliked Location  & Majors            \\
		Preferred Major    & Ranking           \\
		Disliked Major     & Enrollment       \\
		\hline
		\end{tabular}
	\end{table}

\begin{figure}
    \centering
    \captionsetup{justification=centering}
    \includegraphics[width=8cm, height=12cm]{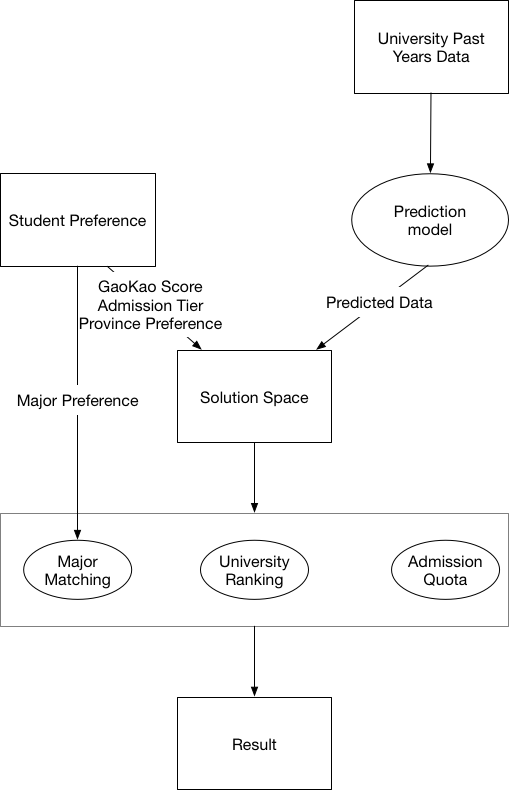}
    \caption{The architecture of the Gaokao application recommendation system}\label{fig:1}
	\end{figure}

    Next, we form a solution space based on the predefined parameters. For each university $U$, find the admission score $A^l_{\ell-1}(U)$ and the highest admission score $A^h_{\ell-1}(U)$. Choose one of the prediction models and use it to predict $A^l_{\ell}(U)$ and $A^h_{\ell}(U)$. %Let~$H_u$~and~$L_u$~denote, respectively, the highest and lowest admission score.
We divide $$D=A^h_{\ell}(U) - A^l_{\ell}(U)$$ into $J$ intervals as follows:
\begin{eqnarray*}
&&[A^l_\ell(U) - \delta_U, I_1), [I_1, I_2), \cdots, [I_{J-2}, I_{J-1}), \\
&&[I_{J-1}, A^h_\ell(U)+\delta_U),
\end{eqnarray*}
where $J$ is the number of universities students are allowed to apply which is
predetermined by students' PAR Gaokao office, $\delta_U$ is a predefined small positive integer depending on $U$ (e.g., we may select $\delta_U = 5$), and
$$I_j = A^l_\ell(U) + j \cdot \frac{D}{J},$$ $j = 1, 2,\ldots, J$. A university is in the solution space if the applicant's Gaokao score falls into one of the intervals. We categorize the solution space
based on the parallel admission model.
For instance, suppose in a parallel admission model, students are allowed to
enter three universities, then $J = 3$ and we have A-, B-, and C-recommendations.
Then the intervals are
\begin{eqnarray*}
A&=&[A^l_\ell(U) - \delta_U, I_1), \\
B&=& [I_1, I_2), \\
C&=& [I_2, A^h_\ell(U)+\delta_U).
\end{eqnarray*}
        A university is in an applicant's A-recommendation when the applicant's Gaokao score falls in the A-interval. Universities listed in the A-recommendation are competitive for the applicant, but the applicant still has a small chance to be admitted. Universities listed in
the B-recommendation present a good match with the applicant, who has a good chance to be accepted. Universities listed in the C-recommendation are the safest choices for the applicants, yet match the applicant's ability. The architecture of the recommendation
system is shown in Figure \ref{fig:1}. The reader is referred to \cite{Lu2016} for more information about the EEZY Gaokao application recommendation system.

\begin{comment}
    Finally, instead of just finding out the universities that match the applicant's preference, we use Base-line Recommendation Index(BRI) to quantify the matching degree. BRI contains three measurements: university ranking, major matching, admission quota. We use the same mechanism we described in EEZY system to compute BRI. To measure the university ranking, we categorized all universities in mainland China into eight groups. For each school, combined the group with the x-recommendation category to compute the university ranking score. To measure the major matching, we devised a mechanism to calculate the similarity between majors based on a hierarchical academic disciplines structure (In China, academic disciplines are officially classified into a hierarchy of three classes). To measure the admission quota, we considered both university enrollment number and the largest enrollment number in the same university group.

    Recommendation result which contains best matching universities categorized by X-recommendations model. Applicants select one university for each category and fill in the application form.
\end{comment}

\section{Experiments and Results}

We consider the following seven features in our experiments, divided into two groups.
The first group consists of the basic information: (1) PAR, (2) exam type, (3) admission tier, and (4) university name. The second group
consists of (5) the highest score, (6) the admission score, and (7) the enrollment number for each university, which are baseline features for prediction and recommendation.
%The
%last group consists of the median score, the average score, and the standard deviations,
%which are necessary to get the best result.

	\subsection{Dataset}

    Enrollment data which contains each student's enrollment information is the best data to generate all the features we need. In addition to the basic information, a useful enrollment dataset should include each student's Gaokao score, admitted university, and admitted major.
%admission score, and ranking.
%With these data,
We can then generate all the features using statistical methods for each university and each major.

    For most PARs, enrollment data is not easy to find. Instead of using all the feature, our baseline features can be satisfied easily. Each PAR publishes a printed Gaokao Application Official Guide that %after Gaokao exam
    contains previous years' Gaokao data to help applicants make decisions. The baseline features can be easily found in the official guide. We devised a procedure to turn printed books into editable excel files using off-the-shelf OCR software.

\subsection{Preprocessing}

After data is collected, we would need to generate, for each PAR and each admission tier,
a score-ranking table of the previous year; and detect  and remove outliers to ensure prediction quality.

\subsubsection{Score-ranking table generation}
A score-ranking table of the current year (i.e., Year $\ell$) is necessary for our prediction models.
% but also is a must-have data for applicants.
With this table, students can easily find their rankings based on their scores.
Some PARs publish this table at the time when the students are notified of their Gaokao scores before the application begins. However, not all PARs provide this information on time. Some other PARs may only publish part of the table (e.g., they only publish a table with rankings for every five points). We devise methods to deal with
these two scenarios by approximating $SRT_{\ell}$.
% based on previous year's score-ranking table $SRT_ell$.

%\textcolor{red}{
Assume that the ranking data is available every five points. We use the PCHIP interpolation method from MATLAB on the current year's incomplete score-ranking table to generate new data points. Figure \ref{fig:2} shows an example of the interpolation results on
2015 Jiangsu Province's officially announced rankings of scores every five points.
Also displayed is the true score-ranking table of 2015.
We can see that the interpolation results match with the true values very well. In particular,
the maximum error of the interpolated points is only 0.5 points smaller than the
true value.
%
%Figure 2 shows an example that we use this interpolation
%method on Jiangsu Province’s 2015 official announced
%incomplete score-ranking table. The maximum error for the
%estimation is below 0.5 point when compared to the complete
%score-ranking table.
%}

%\textcolor{red}{
Assume that none of the score-ranking data for the current year is available. The following method generates an approximation of $SRT_{\ell}$.
%}

\begin{comment}
\begin{enumerate}[label=\alph*)]
%		\item Let~$SRT_\ell$~denote previous year score-ranking table either generated from %enrollment data or obtained from the internet.
        \item For a given PAR, exam type, and tier, find the highest score $H_\ell$ as the upper bound, and use the ASCL $L_\ell$ for the PAR as the lower bound; these numbers can be quickly found from the Internet.
		\item If the score-ranking data of the current year is incomplete, we use Hermite interpolation method to generate new data points within interval $[L_\ell, H_\ell]$.
		\item Separate the revised $SRT_\ell$ into two parts:
		(1) $[H_\ell-L_\ell/5, H_\ell]$,  which is the higher score segment;
		(2) $[L_\ell, H_\ell-L_\ell/5-1]$, which is the lower score segment.
		\item Apply the polynomial curve fitting method on both segments to find two curves that fit the score-rank data of Year $\ell$. Note that if we only use one curve instead of two, then it may result in deviation in the higher score segment.
		\item Combine the two curves, denoted by $C_\ell$.
		%$~denote the difference between previous year and current year .
		%\item
		\textcolor{red}{Divide $[H_\ell, L_\ell]$ evenly into $D$ small intervals. Shift $C_\ell$ based on $D$ to get a new curve $C_{\ell+1}$.}
		\item Generate data points using $C_{\ell+1}$ to form the score-ranking table
		$SRT_{\ell+1}$.
		%needs an example here
		\end{enumerate}
\end{comment}

%\textcolor{red}{
\begin{enumerate}[label=\alph*), font=\itshape]
%		\item Let~$SRT_\ell$~denote previous year score-ranking table either generated from %enrollment data or obtained from the internet.
        \item For a given PAR, exam type, and tier find the highest score $H_\ell$ as the upper bound, and use the ASCL $L_\ell$ of the PAR as the lower bound; these numbers can be quickly found from the Internet.
		\item If the score-ranking data of the current year is incomplete, we apply the PCHIP interpolation method to generate new data points within the interval $[L_\ell, H_\ell]$ to generate $S_\ell$.
			\begin{figure}[h]
    \centering
    \captionsetup{justification=centering}
    \includegraphics[width=9.5cm, height=6.5cm]{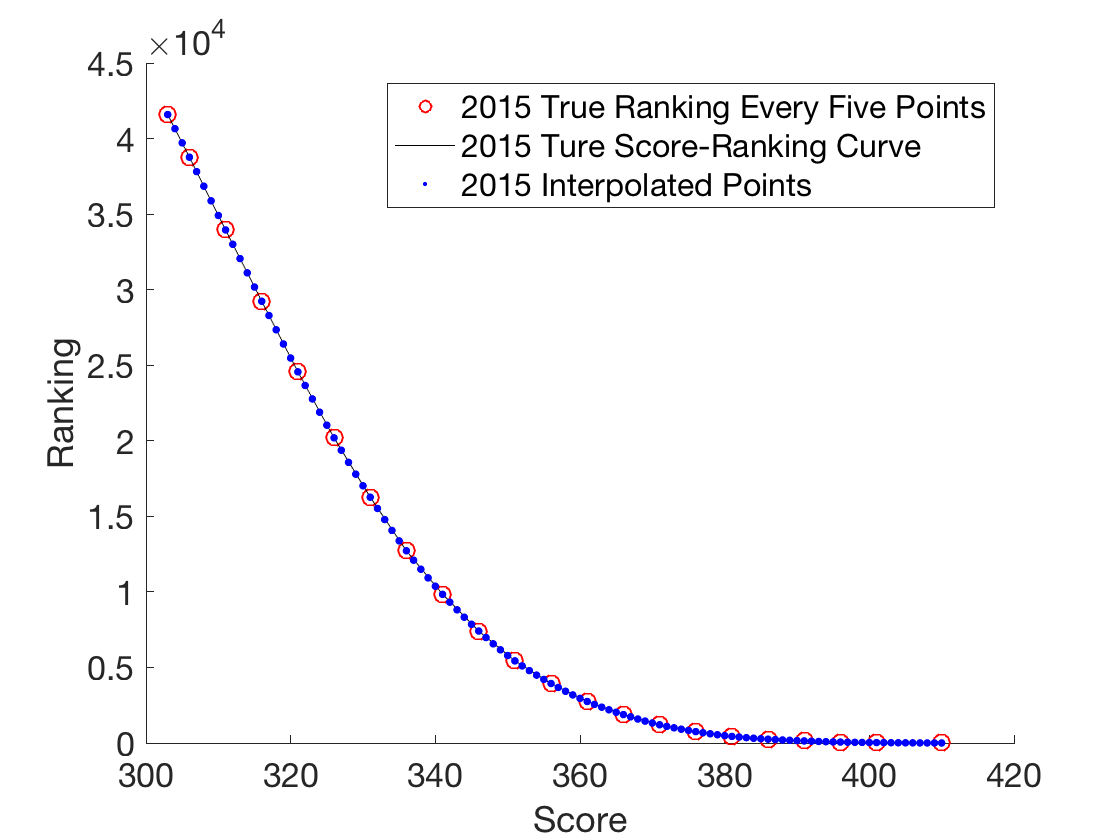}
    \caption{Comparison of the interpolated score-ranking curve with the true score-ranking curve for Jiangsu 2015}\label{fig:2}
	\end{figure}

	\begin{figure}[h]
    \centering
    \captionsetup{justification=centering}
    \includegraphics[width=9.5cm, height=6.5cm]{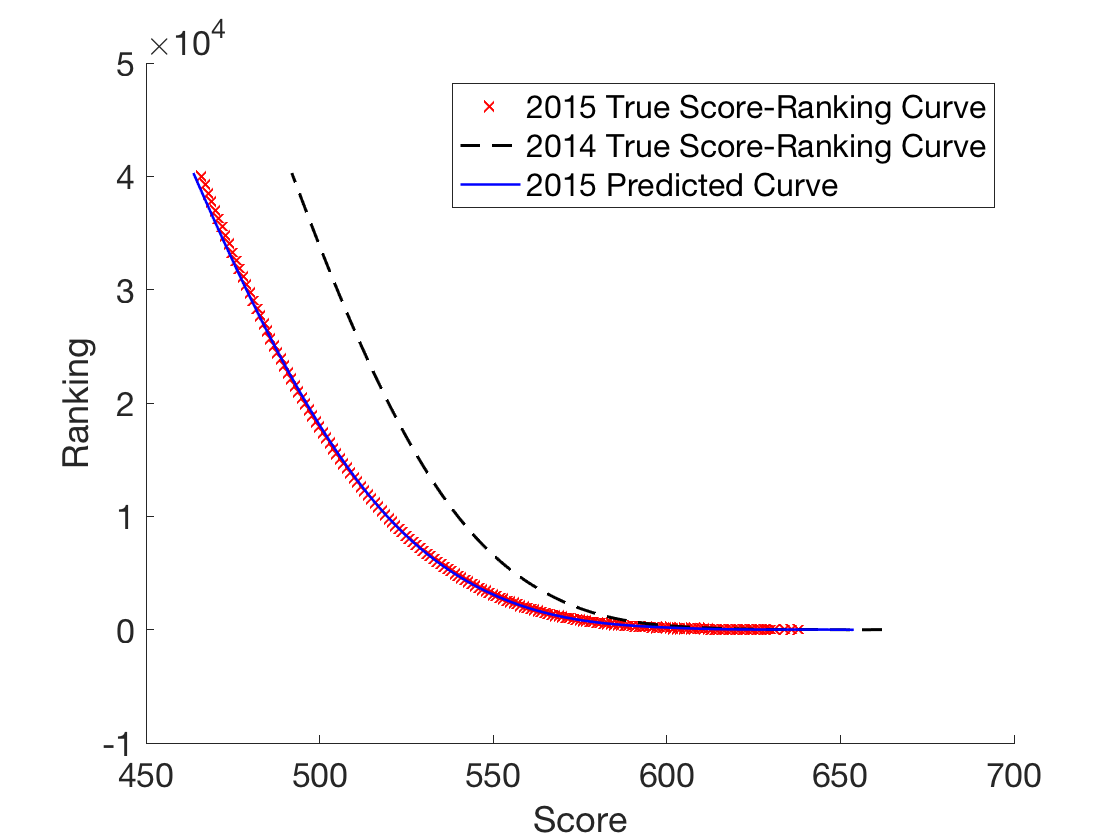}
    \caption{Comparison of the predicted score-ranking curve of Henan 2015  with the true score-to-ranking curve, where the maximum error is 1 point}\label{fig:3}
	\end{figure}

		\item If the score-ranking data for the current year is not available, let $D = L_{\ell} - L_{\ell-1}$.
		For each point $(s_i, r_i)$ in the score-ranking table of Year $\ell-1$, %$S_{\ell-1}$, %score-ranking table
		compute an amended point $(s_i + \delta_i, r_i)$, where
		\[ \delta_i = \left\{
			\begin{array}{ll}
				\bigg\lceil \frac{H_{\ell} - s_i}{H_{\ell-1} - L_{\ell-1}} \times D\bigg\rceil,
				&\mbox{if $D \geq 0$}, \\
				\\
				\bigg\lfloor \frac{H_{\ell-1} - s_i}{H_{\ell-1} - L_{\ell-1}}\times D\bigg\rfloor, &\mbox{otherwise.}
			\end{array}\right.
		\]
		\item Let $RSRT_{\ell-1}$, $RH_{\ell-1}$, and $RL_{\ell-1}$ denote, respectively, the amended score-ranking table, highest score, and lowest score.
		\item Separate the amended $RSRT_{\ell-1}$ into two parts:
		(1) $[RH_{\ell-1}-RL_{\ell-1}/5, RH_{\ell-1}]$,  which is the higher score segment;
		(2) $[RL_{\ell-1}, RH_{\ell-1}-RL_{\ell-1}/5-1]$, which is the lower score segment.
		\item Apply the Fourier curve fitting method on both segments to obtain two curves that fit the score-to-rank data. Note that if we only use one curve instead of two, then it may result in deviation in the higher score segment.
		\item Combine the two curves, denoted by $C_{\ell}$.
		\item Generate data points using $C_{\ell}$ to form the score-ranking table
		$SRT_{\ell}$.
		%needs an example here
		\end{enumerate}

%}
%\textcolor{red}{

Figure \ref{fig:3} shows an example of predicted score-ranking table of 2015 in Henan province using the dataset of 2014, with the maximum error of just 1 point.
%}

\subsubsection{Outlier detection}

The Gaokao policy states that minority nationality students, students from remote rural areas, and students with exceptional talents in sciences or sports may be admitted with lower Gaokao scores. These scores lower the university's admission score, which in turn would affect the prediction results. Thus, it
is necessary to detect and remove such scores and other outliers.
As an example of other outliers, a student of an exceptionally high Gaokao score who could be admitted to
a university of a much higher rank chooses to go to a university of a lower rank because of its location or major.

We use the median absolute deviation (MAD) method \cite{Leys2013} to find outliers.
Sort the student scores in ascending order. Let $S_i$ denote the $i$-th score. That is, $$S_1 \leq S_2 \leq \cdots \leq S_m,$$ where
$m$ is the number of students admitted by $U$ in Year $\ell-1$.

% in university~$U$.
\subsubsection*{Single MAD test}
			\begin{enumerate}[label=\alph*)]
			\item 	Let $M_U$ denote the median of $\{S_1,S_2,\ldots,S_m\}$.
			\item Let $D_i$ denote the difference between a student score and the median. That is,
			$D_i= S_i - M_U$ for $i = 1, 2, \ldots, m$.% compute the difference between student score and median D1,D2... Dm
			\item Let MAD denote the median of $\{D_1, \ldots, D_m\}$. % MAD
			\item For each $S_i$ from both ends, if $$\frac{0.6745 \times |S_i - M_U|}{\mbox{MAD}} \geq 2.24,$$ then $S_i$ is considered an outlier and removed.
			\end{enumerate}

\subsubsection*{Double MAD test}
			\begin{enumerate}[label=\alph*)]
			\item Let
			\begin{eqnarray*}
			S_l &=& \{S_1, \ldots, S_{\lfloor m/2\rfloor}\},\\
			S_h& =& \{S_{\lfloor m/2\rfloor+1},\ldots,S_m\}.
			\end{eqnarray*}
			Let $LM_U$ and $RM_U$ denote, respectively, the median of $S_l$ and $S_h$.
			%score of the first half of the data and the second half of the data. That is, $LM_U$ %is at the first quarter position and $RM_U$ is at the third quarter position.
			\item Let %$D_i$ denote the difference between $S_i$ and the median,
			$D_i = S_i- M_U$ for  $i = 1, 2, \ldots, m$.
			\item Let $\mbox{MAD}_l$ be the median of
			$\{D_1,\ldots, D_{\lfloor m/2\rfloor}\}$
			and $\mbox{MAD}_r$ the median of $\{D_{\lfloor m/2\rfloor}+1,\ldots, D_m\}$.
			\item If $S_i \leq M_U$ and
			$$
			\frac{0.6745 \times |S_i - LM_U|}{\mbox{MAD}_l} \geq 2.24,$$
			then $S_i$ is considered an outlier and removed.

			If $S_i > M_U$ and
			$$\frac{0.6745 \times |S_i - RM_U|}{\mbox{MAD}_r} \geq 2.24,$$
			then $S_i$ is considered an outlier and removed.
			\end{enumerate}

%$S_i$~is an outlier if ~$S_i$~is considered an outlier in single-MAD test or double-MAD %test. Outlier score will be removed and not be used for further computation.

\subsection{Experiment Results}

It is commonly known that certain universities and majors may present the scenario of
``big-year-and-small-year"; namely, in one year for some reasons many students apply to the same major at a certain university, resulting in a much higher admission score than it should be. This discourages students in the next year to apply to this major at this university, resulting in a much lower admission score than it should be. This cycle may repeat, but we argue that it suffices to only look at the most recent two consecutive years of data to
predict admission scores of the new year. In particular, if the reputation of a university's degree programs is stable, then looking at the past two years is sufficient. If it is unstable, either up or down, then looking at  the past two years would be much closer to what it would be today than looking at years further before.

We collected three years (2013--2015) of Gaokao admission data for a number of PARs.
We use the data of 2013 and 2014 for modeling and the data of 2015 for validation.
%the  and used these datasets to test our prediction models.
In this section, we will use the admission data of Henan, Hebei, and Jiangsu provinces to demonstrate our results. Henan and Hebei exams  are
graded using a 750-point grading scale, while the Jiangsu exams are graded using a
480-point grading scale. The accuracies of predictions for all provinces we tested are similar.

We first apply outlier detection method on the dataset to remove outliers.
%and then generate the seven features
%we used for and each year's score-ranking table.
Our models %baseline ranking model, Ranking model, score section model, and Delta model
 do not require training, so we directly run the three models on the 2015 data for evaluating predictions using the 2013 and 2014 data. %\textcolor{red}{
We first apply the 2013 dataset to predict the admission scores of 2015, then do the same using the 2014 dataset.  % The predictive process is separate for each year,
We then compute the mean as the final result to deal with the ``big-year-and-small-year" scenario. %}
 %For average score model, we use 2013 and 2014 year's data for training and use the %parameters we learned to predict 2015 year's data.
We carry out validations separately for each type of exams (the Li-Ke exam and the Wen-Ke exam) and each admission tier. For simplicity, we only consider tier-1 and tier-2 university admissions.

%Unlike other prediction problems, it is notable
Because there are many factors that
would affect the admission score for a particular university, it
is difficult to predict admission scores to be exactly same as the
true admission scores.
%
%that the predicted score same as admission score is nearly impossible even for human experts,
A small error is allowed when experts make predictions.
For an exam of a 750-point grading scale, we would accept errors of
5 to 7 points predicted admission scores, for the reason that
%Therefore, a 7-points difference is allowed in the evaluation, since
such a prediction result is good enough for making robust recommendations. Moreover, the EEZY recommendation model we described in Section 4 can further reduce the impact of such an error using
the parallel admission model.

We also compare our prediction results with the two early models described in Section 2.
They are the % In contrast, we implemented
average admission score model (AASM) and the average admission difference model (AADM).
% described in Section 2 and evaluated them
%   on our dataset. We predict the score as follow: \\
For AASM, the prediction is computed as follows:
$$A_{\ell}(U) = \frac{1}{2}(A_{\ell-1}(U) + A_{\ell-2}(U)).$$
For AADM, the prediction is computed as follows:
$$A_{\ell}(U) =  L_{\ell} + \frac{1}{2}(A_{\ell-1}(U) + A_{\ell-2}(U) - L_{\ell-1}-L_{\ell-2}).$$
As pointed out in Section 2, the current admission only correlates with the past two  years of admission records. These two
models are commonly used by teachers and experts to help students fill out their applications.

Let LK1 denote ``Li-Ke exam for admission tier 1'',
LK2 denote ``Li-Ke exam for admission tier 2'',
WK1 denote ``Wen-Ke exam for admission tier 1'', and
WK2 denote ``Wen-Ke exam for admission tier 2''
Let PD denote ``Point Difference." For example,
by PD = 5 it means that the absolute value of the predicted admission score and
the true admission score is less than or equal to 5.
For each model, the percentage is the number of universities with correctly predicted
admission scores over the total number of universities for a particular tier.
%For example, for WPM, the percentage of 90.6\% for LK1 under PD = 7 means
%that WPM predicts 90.6\% of universities in tier-1 with PD = 7.
%The prediction results are listed in Table \ref{table:3}.
For example, for WPM, the percentage of 90.6\% for LK1 under PD = 7 means
that WPM predicts 90.6\% of tier-1 universities with PD = 7.
The prediction results are listed in Table \ref{table:3}.

\begin{table}
\footnotesize
\centering
\caption{Accuracy of predicting university admission score on data of Henan province by various models}
\label{table:3}
\begin{tabular}{l|c|c|c|c|c}
\hline
\textbf{Model}  & \textbf{PD} & \textbf{LK1 (\%)} & \textbf{WK1 (\%)} & \textbf{LK2 (\%)}  & \textbf{WK2 (\%)}	\\\hline
AASM & \multirow{5}{*}{5} & 12 &41.6 &29.9 &11.9 \\
AADM & &12.8 &21.9 &25.5 &33 \\
BRM & &73.3 &74.1 &64.3 &64.2 \\
WSM & &\textbf{80.5} &76.3 &\textbf{75} &\textbf{70.5} \\
WPM & &79.2 &\textbf{77} &74.2 &68.1 \\
\hline
AASM & \multirow{5}{*}{6} & 15 &47.9 &36.1 &14.7 \\
AADM & &15.3 &24.7 &31 &36.6 \\
BRM & &78.1 &81 &73.2 &70.2 \\
WSM & &\textbf{84.7} &82.5 &\textbf{79.4} &\textbf{76.7} \\
WPM & &84.3 &\textbf{83.6} &78.1 &74.6 \\
\hline
AASM & \multirow{5}{*}{7} & 17.8 &54.3 &42 &19.7 \\
AADM & &20.1 &29.6 &36.3 &44.8 \\
BRM & &84.8 &85.4 &76.3 &77.7 \\
WSM & &89.4 &\textbf{88.7} &82.2 &\textbf{81.1} \\
WPM & &\textbf{90.6} &87.8 &\textbf{83} &80.3 \\
\hline
\end{tabular}
\end{table}

From Table \ref{table:3} we can see that the predictions by the commonly used methods of AASM and AADM are far below satisfaction. We note that, in general,
predictions for tier-1 universities by the three ranking based models are
better for tier-2 universities. This is expected, for the reputations of tier-2 universities are not as stable as the reputations of tier-1 universities.
Besides, the highest accuracy
is produced by WPM, which is close to 91\%.

%\textcolor{red}{In order
%To fully test our model, we evaluated them on Jiangsu and Hebei provinces' data and %achieve similar result.
%Unlike Henan and Hebei which use 750-point grading scale,
Since Jiangsu exams are graded using a 480-point scale, we narrow the acceptable errors range from 5 to 7 to 3 to 5. As Table \ref{table:4} and \ref{table:5} shows, our models all outperform AASM and AADM in Jiangsu and Hebei. The average accuracy for tier-1 universities is typically above 85\% and can be as high as 90.5\%. Moreover, WPM on average produces better predictions than WSM.
%}

\begin{table}
\footnotesize
\centering
\caption{Accuracy of predicting university admission score on data of Jiangsu province by various models}
\label{table:4}
\begin{tabular}{l|c|c|c|c|c}
\hline
\textbf{Model}  & \textbf{PD} & \textbf{LK1 (\%)} & \textbf{WK1 (\%)} & \textbf{LK2 (\%)}  & \textbf{WK2 (\%)}	\\\hline
AASM & \multirow{5}{*}{3} 	&40.3 	&2		&51		&2.6	\\
AADM & 						&55.9 	&61.2	&52.4	&60.9		\\
BRM & 						&66.8 	&60.7	&58.1	&57.9	\\
WSM & 						&\textbf{74.4}	&74.4		&59.8	&63.9	\\
WPM & 						&73		&\textbf{81.3}	&\textbf{61.4} &\textbf{67.6}	\\
\hline
AASM & \multirow{5}{*}{4} 	&51.2	&3.9		&56.4	&3.4 \\
AADM & 						&63		&73.7	&55.7	&71.1 \\
BRM & 						&76.2	&70.7	&65.1	&70.7 \\
WSM & 						&79.2	&80.7	&66.3	&\textbf{74.8} \\
WPM & 						&\textbf{80.6} 	&\textbf{84.1}	&\textbf{69.8}	&74.5 \\
\hline
AASM & \multirow{5}{*}{5} 	&61.6	&5.3 	&62.6	&3.7 \\
AADM & 						&65.4	&74.3 	&63.5	&74.3 \\
BRM & 						&82		&82 		&71.9	&76.8 \\
WSM & 						&\textbf{85.8}	&84.7 	&72.3 	&82.2 \\
WPM & 						&84.8	&\textbf{90.5} 	&\textbf{75.4} 	&\textbf{83.1} \\
\hline
\end{tabular}
\end{table}

\begin{table}
\footnotesize
\centering
\caption{Accuracy of predicting university admission score on data of Hebei province by various models}
\label{table:5}
\begin{tabular}{l|c|c|c|c|c}
\hline
\textbf{Model}  & \textbf{PD} & \textbf{LK1 (\%)} & \textbf{WK1 (\%)} & \textbf{LK2 (\%)}  & \textbf{WK2 (\%)}	\\\hline
AASM & \multirow{5}{*}{5} 	&24.9 	&44.7	&37.3	&29.1	\\
AADM & 						&22.2 	&9		&8.4		&27.8		\\
BRM & 						&62.3 	&73.2	&54.8	&57.7	\\
WSM & 						&70		&74.8	&\textbf{62.9}	&60.6	\\
WPM & 						&\textbf{71.6}	&\textbf{76.3}	&57.3 	&\textbf{61.6}	\\
\hline
AASM & \multirow{5}{*}{6} 	&27.2	&52.3	&43.8	&34.9 \\
AADM & 						&24.5	&13.1	&9.8		&34.4 \\
BRM & 						&69.7	&78.8	&62.3	&64 \\
WSM & 						&\textbf{77.8}	&81.3	&\textbf{71.2}	&\textbf{73.5} \\
WPM & 						&76.3 	&\textbf{83.8}	&70.1	&73 \\
\hline
AASM & \multirow{5}{*}{7} 	&33.9	&55.8 	&49.9	&40.7 \\
AADM & 						&28		&15.1 	&11.2	&40.7 \\
BRM & 						&77.8	&82.8 	&68.7	&69.6 \\
WSM & 						&82.9	&\textbf{86.4} 	&\textbf{74.2} 	&77.2 \\
WPM & 						&\textbf{84.1}	&85.9  	&\textbf{74.2} 	&\textbf{80.2} \\
\hline
\end{tabular}
\end{table}

\textbf{Remark}.
Further improvement of predictions can be obtained. For example, we note that
the Gaokao scores of admitted students for a university follow a certain distribution close to
being Gaussian, which allows us to extract additional features to help devise prediction models. We can also use supervised machine learning methods to train
prediction models. Using these techniques, we can obtain better predictions.
In particular, our preliminary results show that a 94\% prediction accuracy for tier-1 universities can be achieved.
We plan to report this work on a separate paper.

\section{Conclusions and Future Work}

We presented three models for predicting university admission scores in mainland China and evaluated these methods through extensive experiments. Our models achieved a considerably good prediction accuracy on the testing dataset. Moreover,
we modified an existing
recommendation system
using morphological analysis.
%
%has been proposed to help applicants find the best matching universities.
For future work, we will explore other prediction methods to improve accuracy. For example, we may use sequential data for a multi-year model. We are also interested in additional parameters to
improve recommendation, including annual family incomes, occupations of parents, rankings of majors, perspectives of finding jobs of different majors, among other things.

\section*{Acknowledgment}

This work was supported in part by Massachusetts Education International Inc and Tianhan Technology Inc under research and development grants. We are grateful to Shan Lu, Cheng Zhang, Wenjing Yang, Liqun Shao,
and Xueyuan Guo for their help.

\ifCLASSOPTIONcaptionsoff
  \newpage
\fi



%\bibliographystyle{IEEEtran}

%\begin{biography}[{\includegraphics[width=1in,height=1.25in,clip,keepaspectratio]{mshell}}]{Michael Shell}

\begin{IEEEbiography}[{\includegraphics[width=1in,height=1.25in,clip,keepaspectratio]{picture}}]{John Doe}
\blindtext
\end{IEEEbiography}


%\vfill


\end{document}